\documentclass{osa-article}

\journal{osac}

\articletype{Research Article}
\begin{document}

\title{Direct observation of the spin-orbit coupling interaction in ring-core optical fibers}

\author{Steven D. Johnson,\authormark{1,*}, Zelin Ma,\authormark{2}, Miles Padgett,\authormark{1} and Siddharth Ramachandran\authormark{2}}

\address{\authormark{1} SUPA, University of Glasgow, School of Physics and Astronomy, Glasgow, United Kingdom\\
\authormark{2} Boston University, 8 St. Mary's Street, Boston, Massachusetts, USA}

\email{\authormark{*}steven.johnson@glasgow.ac.uk} 


\begin{abstract}
Ring-core optical fibers have been designed to carry orbital angular momentum modes. We demonstrate the imaging of these modes, individually identifying modes separated temporally by only 30~ps. A single-pixel camera operating in the short-wave infrared detection range is used to image the 1550~nm wavelength optical modes. With this technique, examination of these optical modes can be performed with significantly higher temporal resolution than is possible with conventional imaging systems, such that the imaging of modes separated by spin-orbit coupling is achieved and evaluated. Deconvolution is required to separate the instrument response from the optical mode signal, increasing the clarity and temporal resolution of the measurement system.
\end{abstract}

\section{Introduction}
There is an ever-increasing demand on the transfer of data through communications infrastructure, this increasing transmission of data around the world places greater demand on optical fiber communications. The majority of these optical communications is via single-mode optical fibers, however higher data-rates can be transmitted by using multimode fibers \cite{Essiambre2012}. To respond to this demand for data, multimode fiber systems have been studied \cite{Ryf2015_MMFTransmssion, Lengle2016_MMFTrans}. One proposed method to transmit higher data-rates is to use different orbital angular momentum (OAM) modes, multiplexed into channels in this basis \cite{Ramachandran2013vortices, Bozinovic2013terabit}. Within these structured fibers there is mode dispersion, as the optical modes travel at different group velocities within the optical fiber. To better understand the propagation of the OAM modes within these specialised optical fibers the output needs to be imaged. Here we demonstrate the observation of OAM modes and the interaction of these modes to form interference patterns at the output facet of the optical fiber. To measure these optical modes a single-pixel camera sensitive in the short-wave infra-red (SWIR) is used, with sufficiently high temporal resolution to image the modes separated by spin-orbit coupling within the optical fiber. 

The OAM modes are characterised by the helicity of the wavefront \cite{Allen1992}, where the OAM is expresses as an integer value $L$. These modes have a phase singularity at the centre of the beam such that there is a ring of intensity. The OAM is defined to be positive for counter-clockwise phase rotation and negative for clockwise phase rotation. In addition, the polarisation direction for circularly polarised light can also be in either direction. The group velocity of the mode propagation is dependent on whether the OAM is aligned or anti-aligned to the polarisation direction \cite{Gregg2015}. Not only do the different OAM modes travel with different group velocity, but there is a splitting within these modes due to a spin-orbit coupling within the optical fiber. Therefore, there is a delay between the modes where the spin-orbit coupling is aligned (SOa) or anti-aligned (SOaa).

These small differences in group velocity can only be measured by using many kilometres of optical fiber, where the modes can become separated out by tens of nanoseconds, or by using a camera with very high temporal resolution to image these optical modes. There are a number of options for performing fast imaging: such as a streak camera, SPAD array or gated-intensifier \cite{Faccio2018}. At the telecommunications wavelength 1550~nm a camera with a silicon detector would not be sensitive to this wavelength. Pixelated array-detectors have been produced with InGaAs detectors, but they are of limited used for this measurement due to their low frame rate \cite{Bradley2019_3DInGaAs}. Previous techniques to measure optical modes have used cross-correlation C$^2$ or S$^2$ \cite{Schimpf2011, Demas2014SubSecondC2, Nicholson2008}. Here we demonstrate the imaging of optical modes with the temporal spacing of several picoseconds using a camera system with very high temporal resolution.

The alternative to using a conventional array-detector is to sample the image part-by-part, measuring signal from areas of the image frame; this is made possible by using a transformable mask \cite{Edgar2019SinglePixel}. This technique uses a single-element detector to construct the whole image, known as single-pixel imaging. As the single-pixel imaging requires multiple measurements to be taken it cannot be performed as a single-shot method, which is similar to other methods used to image fast phenomena \cite{Gariepy2015}. Using a single-pixel camera and recording the time signal measured for each mask the image at a series of time periods can be calculated, this is effectively a high-speed video of the captured light. This method of using a single-pixel camera builds upon previous work measuring optical fiber modes \cite{Johnson2019} where the method was demonstrated with an order of magnitude lower temporal resolution, lower signal-to-noise and longer acquisition times. 



\section{Method}
\begin{figure*}[tbp]
    \centering
    \includegraphics[width=0.95\textwidth]{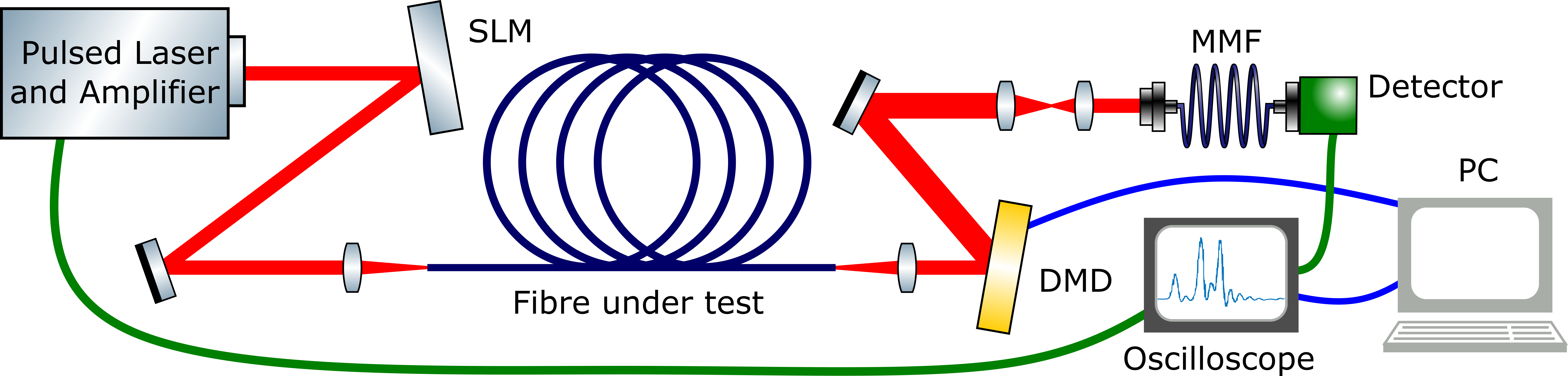} 
    \caption{{\bf The layout of the optical fiber and single-pixel camera}. The pulsed laser is incident on a spatial light modulator (SLM) to precisely adjust the coupling into the optical fiber. The light is coupled into and travels through the ring-core fiber, and the collimated output is incident on a digital micromirror device (DMD). The signal reflected by each mask displayed on the DMD is measured by a fiber-coupled high-speed photodetector via a multimode fiber (MMF). The signal is measured by an oscilloscope triggered from the laser; data is transferred to the PC for analysis.}
    \label{fig:SetUp}
\end{figure*}

The layout of the experiment is shown in figure \ref{fig:SetUp}. The pulsed laser used was a PriTel FFL Series Picosecond Fiber Lasers at a wavelength of 1550~nm, with a repetition rate of 20~MHz and pulse width of 2~ps and timing jitter of <1~ps. To increase the laser power a PriTel PMFA Series Erbium doped fiber amplifier was used. A 1550~nm bandpass filter removed any unwanted light from the amplified spontaneous emission. To control the light coupled into the optical fiber under test a Hamamatsu Liquid Crystal on Silicon Spatial Light Modulator (SLM) was used to apply a ring-pattern phase to the beam, this beam was imaged onto the facet of the fiber under test. We used a MEMS device known as a digital micromirror device (DMD) to produce our transformable mask. The far-field of the optical fiber facet was imaged on to the ViALUX V-7001 DMD, which had $1024 \times 768$ mirrors with a 13.7~{\textmu}m pitch. Due to the small beam size only a small subset of the mirrors was used for the measurements. The light selected by the DMD was coupled into a 50~{\textmu}m core multimode optical fiber (MMF) that connected directly to a Newport 1444-50 fiber-coupled high-speed photodetector, this time domain optimised photodetector had an 18.5~ns impulse response. The temporal measurement was performed with an 86100A Infiniium DCA Wide-Bandwidth Oscilloscope, the 50 GHz oscilloscope recorded the temporal signal for each pattern displayed on the DMD, with respect to a trigger signal from the picosecond laser. The measurement was transferred to the PC, which calculated the temporally changing pattern of the optical modes. 

For ease of use an orthogonal set of binary patterns, the Hadamard pattern set, were used as the mask set \cite{Pratt1969}. An alternative method would be to sample pixel-by-pixel, the downside of this method is the reduced integration time for each pixel in the resulting image, with this scheme each pixel is only measured once and hence there is much lower signal-to-noise in the subsequent measurement. The Hadamard pattern set comprises of elements taking the value of +1 or -1, as a DMD can only modulate the intensity on or off two measurements are made. The +1 values are displayed on the DMD and a measurement is taken, then repeated with the -1 values.

For each mask $i$ the temporal signal is measured on the oscilloscope. For all points in the temporal signal $M(t)_i$ is calculated as the subtraction of the signals for the positive and negative patterns. The spatial image $I_\text{2D}$ is calculated via a Hadamard transform using the signal and Hadamard pattern matrix $H_i$. The image is reconstructed as
\begin{equation}
    I_\text{2D}(t) = \sum_{i=1}^{N} M(t)_{i} \cdot H_{i},
    \label{Eq1}
\end{equation}
where $N$ is the number of Hadamard patterns, also corresponding to the number of pixels in the reconstructed image. The 2D intensity varying over time $t$ produces a 3D data set of the changing mode picture. 

\begin{figure*}[tbp]
    \centering
    \includegraphics[width=0.95\textwidth]{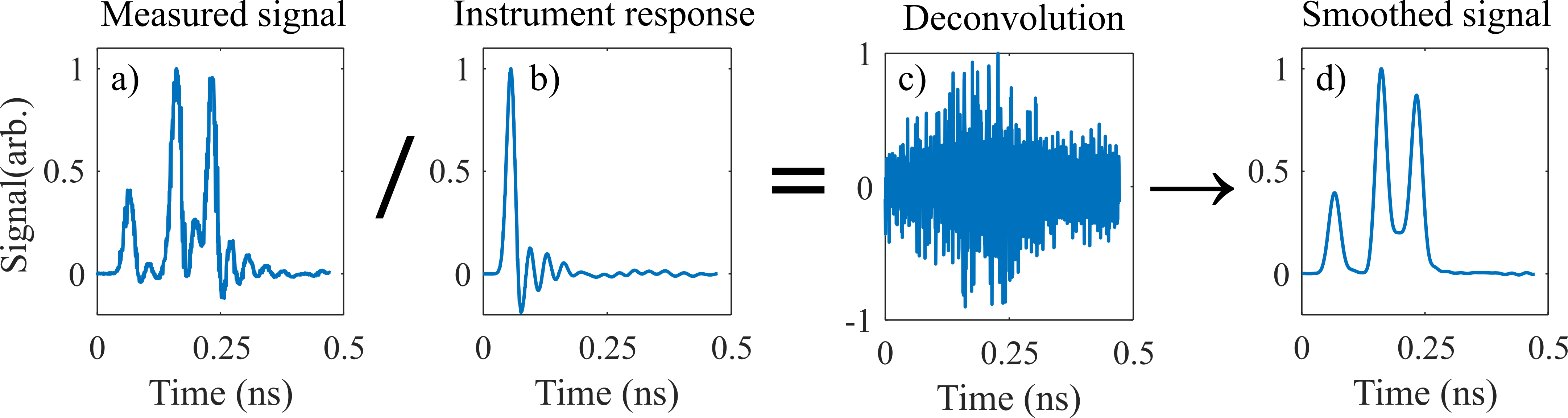}
    \caption{{\bf Example of the deconvolution process.} Using the data set presented later in figure \ref{fig:MainResult}c) an example of the deconvolution is shown. The raw data {\bf a)} is deconvolved with the measured impulse response {\bf b)}, producing the resulting signal {\bf c)}. The signal is then Gaussian smoothed to produce the final signal {\bf d)}.}
    \label{fig:Deconvo}
\end{figure*}

Due to the ringing in the impulse response of the detector, the raw measurements required deconvolution to enable the individual modes to be identified, as illustrated in figure \ref{fig:Deconvo}. The impulse response was measured separately to the system by using an attenuated signal directly from the laser source. To deconvolve the measured signal Fourier deconvolution was used, followed by a Gaussian smoothing that produced the signal used for reconstruction. The deconvolution was performed after the individual pixel values were calculated from the Hadamard transform instead of performing on deconvolution on the raw temporal data, although there was little difference in the results when either method was used. 

\section{Results}

\begin{figure*}[tbp]
    \centering
    \includegraphics[width=0.90 \textwidth]{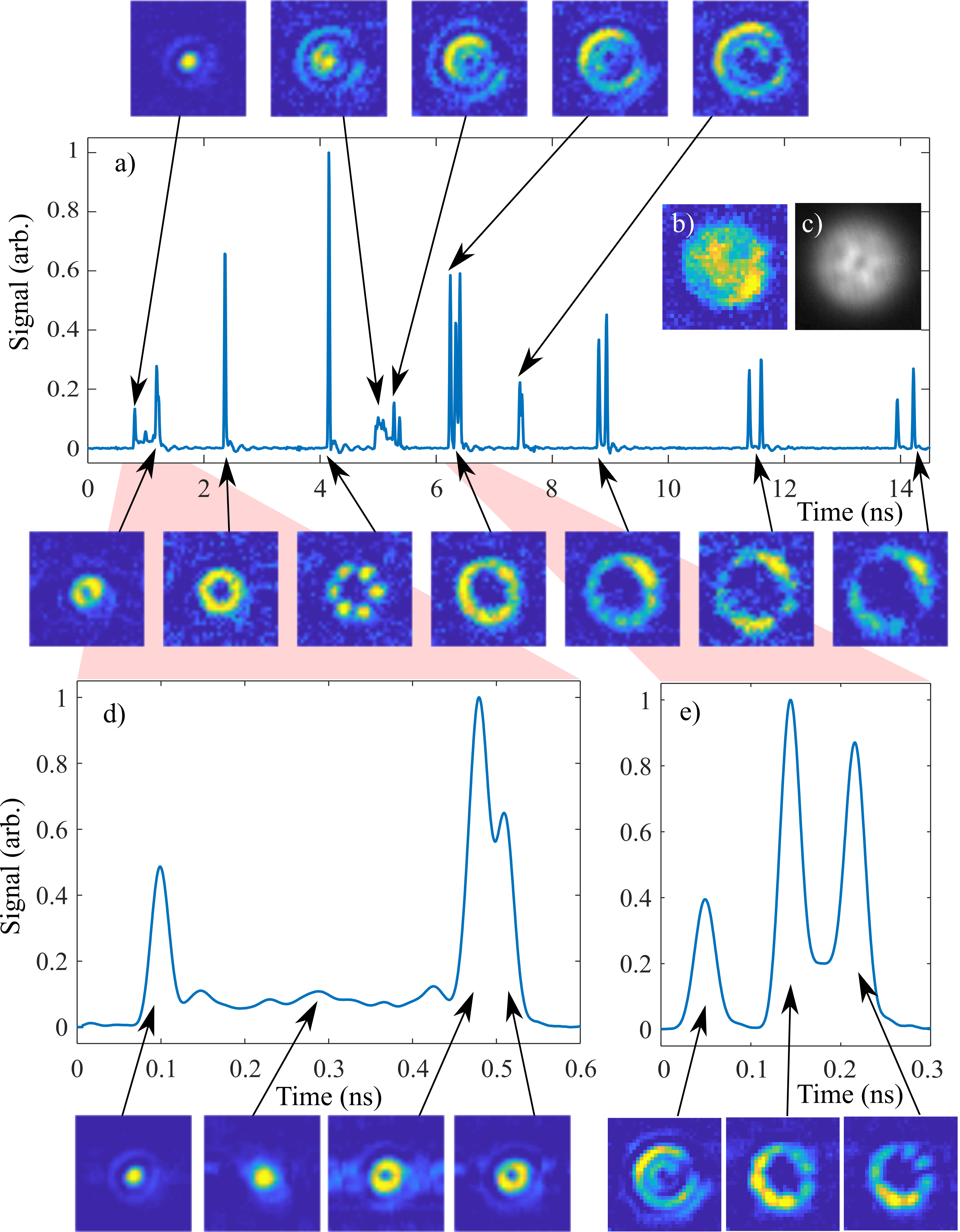} 
    \caption{{\bf Measurement of the optical modes}. Each data set is shown with the total signal measured over time and the modes present for the peaks indicated by the arrows. {\bf a)}~shows the full measurement of all present modes. The modes associated with single OAM are shown below the time series (modes $L=\pm1$ to $L=\pm7$), other modes are shown above. The image {\bf b)} is the summed signal for all time bins, showing similar signal to {\bf c)}, an image recorded by a Goldeye P-008 SWIR camera with an InGaAs sensor. {\bf d)}~shows the first modes where there is strong coupling between the fundamental Gaussian and SOa$_{0,1}$ and SOaa$_{0,1}$ modes. {\bf e)}~shows the SOa$_{3,2}$/SOaa$_{3,2}$ and SOa$_{4,1}$/SOaa$_{4,1}$ modes. The interference of the $+L$ and $-L$ modes produce the petal pattern as shown in the image. 
    }
    \label{fig:MainResult}
\end{figure*}

The optical fiber under test was a ring-core fiber presented in the work by Gregg et. al. \cite{Gregg2015}, the fiber core had a radius of 8.25~{\textmu}m with an air-core of 3~{\textmu}m and the length of fiber used was approximately 200~m. All measurements presented were for a $32\times32$ pixel image, the positive and negative patterns were shown for each Hadamard pattern requiring a total of 2048 patterns per acquisition to be shown sequentially on the DMD. For each acquisition set 1800 points were taken across the desired period from a single scan on the oscilloscope for each Hadamard pattern displayed. The total acquisition time required was around 20 minutes, this could become much faster with triggering between the DMD pattern display and the temporal measurement as well as faster transfer of the data from the oscilloscope. 

The coupling position and beam shape produced by the SLM were adjusted to couple similar signal levels to each of the optical modes, these were adjusted to maximise the power for each acquisition. The data presented are from three data acquisitions, one showing the measurement of all modes supported by the optical fiber (figure \ref{fig:MainResult}a) and two with higher temporal resolution focusing on a period of interest (figure \ref{fig:MainResult}b and \ref{fig:MainResult}c). The full measurement was taken for a 15~ns acquisition. The fiber carries the OAM modes with the expected ring-modes, these are observed to be separated into the SOa and SOaa modes when the time delay difference is greater than the detector temporal resolution. For our system there is no polarisation control of the input light, such that both aligned and anti-aligned modes are present within the optical fiber. If there is interference of these two components the interference lobes are seen in the measured intensity. This petal pattern is present in many of the modes, where the number of petals is twice the order of orbital angular momentum order $L$. 

The modes that are seen in figure \ref{fig:MainResult}a) show the fundamental OAM modes below the plot, these are (from left to right) the SOa$_{1,1}$/TE$_{0,1}$/TM$_{0,1}$, SOa$_{2,1}$/SOaa$_{2,1}$, SOa$_{3,1}$/SOaa$_{3,1}$, SOa$_{4,1}$/SOaa$_{4,1}$, SOa$_{5,1}$/SOaa$_{5,1}$, SOa$_{6,1}$/SOaa$_{6,1}$ and SOa$_{7,1}$/SOaa$_{7,1}$ modes. Above the plot are the modes without OAM and the OAM modes with higher radial order, which are (from left to right) the SOa$_{0,1}$, SOa$_{0,2}$/TE$_{0,2}$/SOa$_{1,2}$/TM$_{0,2}$, SOa$_{2,2}$/SOaa$_{2,2}$, SOa$_{3,2}$/SOaa$_{3,2}$, and SOa$_{4,2}$/SOaa$_{4,2}$ modes. In both cases it is observed that for the increasing OAM order the ring becomes larger in the far field. Some modes show a low haystack (like SOa$_{0,2}$/TE$_{0,2}$/SOa$_{1,2}$/TM$_{0,2}$) instead of a sharp peak, due to the distributed coupling of different modes whose effective indices are very close.

\begin{figure*}[tbp]
    \centering
    \includegraphics[width=0.60 \textwidth]{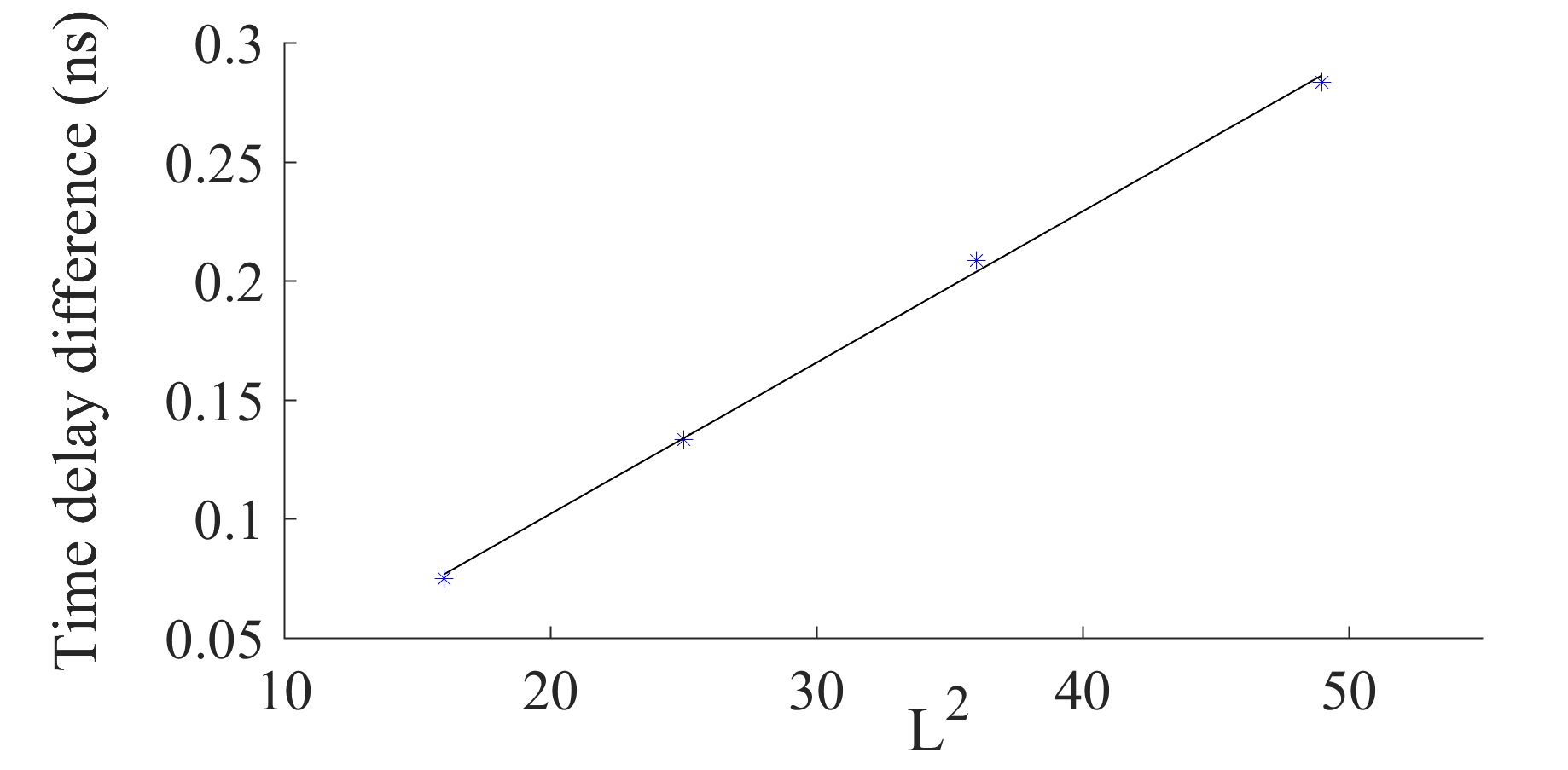}
    \caption{{\bf Difference in arrival times as a function of mode order $L$.} Through the analysis of the propagation times measured in figure \ref{fig:MainResult}a) for the modes $L=\pm4$ to $L=\pm7$, where the time delay between the SOa and SOaa modes is larger than the temporal resolution of our experiment, the relation between the time delay difference and the mode order $L^2$ is shown to be linear.}
    \label{fig:L^2_measurement}
\end{figure*}

The measurement provides a direction observation of the spin-orbit coupling in fibers, which arises from the anisotropy of the fiber's refractive index distribution. Similar to atomic spin-orbit interaction that splits the degeneracy of electronic energy level, the photonics spin-orbit interaction in fiber lifts the degeneracy of propagation constant $\beta$ \cite{Leary2009}. Such splitting of propagation constant between the SOa and SOaa modes $\delta\beta$, known to enabled stable OAM propagation in fibers \cite{Gregg2015,Ramachandran2009}, can be written as
\begin{equation}
    \delta\beta=\frac{L}{ka^2n_{co}^2}\int_{0}^{\infty}{\left|E(r)\right|^2\frac{\partial \Delta n(r)}{\partial r}dr}
\end{equation}
where $r$ is the radial coordinate, $E(r)$ is the normalized electric field for the unperturbed mode, $a$ is the size of the fiber core, $k$ is the free-space wave vector defined as $k=2\pi/\lambda$, $\Delta n(r)$ is the refractive index profile relative to the index of the infinite cladding, and $n_{co}$ is the maximum refractive index. The integral, known as the polarization correction integral \cite{Brues2008_GuidedOptics,Snyder1983_OpticalWaveguide}, depends on the interaction of the field with index gradients of the waveguide, and is typically of the order $L$ in fibers with core sizes <16~{\textmu}m. Thus, the overall effect of OAM order $L$ on the splitting of the propagation constants (as well as group velocities) should scale as $L^2$. Figure \ref{fig:L^2_measurement} is a plot of the difference in arrival times, as a function of mode order $L$, for the modes whose temporal separation was measurable in our experiment. The excellent match with a straight line fit (linear regression R$^2$ coefficient 0.999) of the data clearly indicates that the modes in this fiber follow the splitting as dictated by the spin-orbit interaction of light and the associated polarization correction.

The higher detail measurements (figure \ref{fig:MainResult}b and \ref{fig:MainResult}c) were taken using the same 1800 points but spread across a 1~ns acquisition, focusing on sets of optical modes. These show a clearer measurement of the optical modes as there is longer period of measurement within each time domain. There are modes shown in figure \ref{fig:MainResult}c where the measured optical modes are separated by 30~ps, demonstrating the possible measurement limits of this single-pixel measurement system. For this camera the absolute temporal limit would be due to the detector, although the response time is 18~ps it would be expected a 10~ps difference could be detected. 

The fiber-coupled sensor does place limitations on the size of mode that can be imaged, due to the numerical aperture of the MMF coupled to the detector. Within the system there is significant loss due to the DMD, it is designed for visible wavelengths and for our measurement a maximum of only 10\% of the incident light is reflected from the device. The system is straightforward to align, with coupling into the fiber under test requiring the most precision to produce measurements with evenly distributed optical power into the range of modes. 

\section{Discussion} 
The ability to transmit high data-rates through multimode optical fibers makes this research important to understand mode propagation. Using a single-pixel camera high temporal resolution measurements of the modes have been produced demonstrating an order of magnitude improvement in temporal resolution, lower signal-to-noise and shorter acquisition times than previous measurements \cite{Johnson2019}. The previous work used single-photon counting detectors requiring longer acquisition time to produce an image, analogue detectors have therefore produced much better results. This method of imaging the optical modes has be demonstrated to study the spin-orbit coupling within the optical fiber and shown to match a model of propagation for spin-orbit modes. Further measurements with these optical fibers could be performed to measure the interaction of the spin-orbit aligned and anti-aligned modes by adding a polarisation measurement. The system could also be applied to other classes of multimode optical fibers to understand the mode propagation in a range of systems.

\section{Funding}
Engineering and Physical Sciences Research Council (EPSRC) QuantIC (EP/M01326X/1); H2020 European Research Council (ERC) (TWISTS, 192382) (PhotUntangle, 804626); Office of Naval Research (ONR) MURI (N0014-13-1-0672); National Science Foundation (NSF) (ECCS-1610190).

\section{Acknowledgments}
S.D.J. acknowledges financial support from Scottish Universities Physics Alliance (SUPA) travel grant and the UK Quantum Imaging Hub (QUANTIC).

\section{Disclosures}
The authors declare that there are no conflicts of interest related to this article.

\bibliography{sample}

\begin{thebibliography}{10}
\newcommand{\enquote}[1]{``#1''}

\bibitem{Essiambre2012}
R.-J. Essiambre and R.~W. Tkach, \enquote{Capacity trends and limits of optical
  communication networks,} {\protect\JournalTitle{Proceedings of the IEEE}}
  \textbf{100}, 1035--1055 (2012).

\bibitem{Ryf2015_MMFTransmssion}
R.~Ryf, N.~K. Fontaine, H.~Chen, B.~Guan, B.~Huang, M.~Esmaeelpour, A.~H.
  Gnauck, S.~Randel, S.~Yoo, A.~Koonen, R.~Shubochkin, Y.~Sun, and R.~Lingle,
  \enquote{Mode-multiplexed transmission over conventional graded-index
  multimode fibers,} {\protect\JournalTitle{Opt. Express}} \textbf{23},
  235--246 (2015).

\bibitem{Lengle2016_MMFTrans}
K.~Lengl\'{e}, X.~Insou, P.~Jian, N.~Barr\'{e}, B.~Denolle, L.~Bramerie, and
  G.~Labroille, \enquote{4x10 {Gbit/s} bidirectional transmission over 2 km of
  conventional graded-index {OM1} multimode fiber using mode group division
  multiplexing,} {\protect\JournalTitle{Opt. Express}} \textbf{24},
  28594--28605 (2016).

\bibitem{Ramachandran2013vortices}
S.~Ramachandran and P.~Kristensen, \enquote{Optical vortices in fiber,}
  {\protect\JournalTitle{Nanophotonics}} \textbf{2}, 455--474 (2013).

\bibitem{Bozinovic2013terabit}
N.~Bozinovic, Y.~Yue, Y.~Ren, M.~Tur, P.~Kristensen, H.~Huang, A.~E. Willner,
  and S.~Ramachandran, \enquote{Terabit-scale orbital angular momentum mode
  division multiplexing in fibers,} {\protect\JournalTitle{Science}}
  \textbf{340}, 1545--1548 (2013).

\bibitem{Allen1992}
L.~Allen, M.~W. Beijersbergen, R.~J.~C. Spreeuw, and J.~P. Woerdman,
  \enquote{Orbital angular momentum of light and the transformation of
  {Laguerre-Gaussian} laser modes,} {\protect\JournalTitle{Phys. Rev. A}}
  \textbf{45}, 8185--8189 (1992).

\bibitem{Gregg2015}
P.~Gregg, P.~Kristensen, and S.~Ramachandran, \enquote{Conservation of orbital
  angular momentum in air-core optical fibers,} {\protect\JournalTitle{Optica}}
  \textbf{2}, 267--270 (2015).

\bibitem{Faccio2018}
D.~Faccio and A.~Velten, \enquote{A trillion frames per second: the techniques
  and applications of light-in-flight photography.}
  {\protect\JournalTitle{Reports on Progress in Physics}} \textbf{81}, 105901
  (2018).

\bibitem{Bradley2019_3DInGaAs}
C.~P. Bradley, S.~S. Mukherjee, A.~D. Reinhardt, P.~F. McManamon, A.~O. Lee,
  and V.~Dhulla, \enquote{{3D} imaging with $128\times128$ eye safe {InGaAs}
  p-i-n lidar camera,} {\protect\JournalTitle{Proc. SPIE 11005, Laser Radar
  Technology and Applications XXIV}}  (2019).

\bibitem{Schimpf2011}
D.~N. Schimpf, R.~A. Barankov, and S.~Ramachandran, \enquote{Cross-correlated
  ({C}2) imaging of fiber and waveguide modes,} {\protect\JournalTitle{Opt.
  Express}} \textbf{19}, 13008--13019 (2011).

\bibitem{Demas2014SubSecondC2}
J.~Demas and S.~Ramachandran, \enquote{Sub-second mode measurement of fibers
  using {C2} imaging,} {\protect\JournalTitle{Optics Express}} \textbf{22},
  23043--23056 (2014).

\bibitem{Nicholson2008}
J.~W. Nicholson, A.~D. Yablon, S.~Ramachandran, and S.~Ghalmi,
  \enquote{Spatially and spectrally resolved imaging of modal content in
  large-mode-area fibers,} {\protect\JournalTitle{Opt. Express}} \textbf{16},
  7233--7243 (2008).

\bibitem{Edgar2019SinglePixel}
M.~P. Edgar, G.~M. Gibson, and M.~J. Padgett, \enquote{Principles and prospects
  for single-pixel imaging,} {\protect\JournalTitle{Nature Photonics}}
  \textbf{13}, 13--20 (2019).

\bibitem{Gariepy2015}
G.~Gariepy, N.~Krstaji\'{c}, R.~Henderson, C.~Li, R.~R. Thomson, G.~S. Buller,
  B.~Heshmat, R.~Raskar, J.~Leach, and D.~Faccio, \enquote{Single-photon
  sensitive light-in-fight imaging,} {\protect\JournalTitle{Nature
  Communications}} \textbf{6}, 6021 (2015).

\bibitem{Johnson2019}
S.~D. Johnson, D.~B. Phillips, Z.~Ma, S.~Ramachandran, and M.~J. Padgett,
  \enquote{A light-in-flight single-pixel camera for use in the visible and
  short-wave infrared,} {\protect\JournalTitle{Opt. Express}} \textbf{27},
  9829--9837 (2019).

\bibitem{Pratt1969}
W.~K. Pratt, J.~Kane, and H.~C. Andrews, \enquote{Hadamard transform image
  coding,} {\protect\JournalTitle{Proceedings of the IEEE}} \textbf{57}, 58--68
  (1969).

\bibitem{Leary2009}
C.~C. Leary, M.~G. Raymer, and S.~J. van Enk, \enquote{Spin and orbital
  rotation of electrons and photons via spin-orbit interaction,}
  {\protect\JournalTitle{Physical Review A}} \textbf{80}, 061804 (2009).

\bibitem{Ramachandran2009}
S.~Ramachandran, P.~Kristensen, and M.~F. Yan, \enquote{Generation and
  propagation of radially polarized beams in optical fibers,}
  {\protect\JournalTitle{Optics Letters}} \textbf{34}, 2525--2527 (2009).

\bibitem{Brues2008_GuidedOptics}
J.~Bures, \emph{Guided Optics: Optical Fibers and All-fiber Components} (Wiley,
  2008).

\bibitem{Snyder1983_OpticalWaveguide}
A.~Snyder and J.~Love, \emph{Optical Waveguide Theory} (Springer, 1983).

\end{thebibliography}

\end{document}